\documentclass[twocolumn,aps,superscriptaddress,showpacs,floatfix,prc]{revtex4}
\usepackage{mathrsfs}
\usepackage{amssymb}
\usepackage{amsmath}
\usepackage{graphicx}
\usepackage[normalem]{ulem}
\usepackage[dvips]{color}
\usepackage{bm}
\usepackage{longtable}
\usepackage{times}

\renewcommand\sout{\bgroup \color{red} \ULdepth=-.5ex \ULset}

\renewcommand{\v}[1]{\textbf{#1}}
\renewcommand{\rm}[1]{\textrm{#1}}
\renewcommand{\d}{\mathrm{d}}

\begin{document}
\title{Proton-skins in momentum space and  neutron-skins in coordinate space in heavy nuclei}

\author{Bao-Jun Cai\footnote{Email: bjcai87@gmail.com}}
\affiliation{Department of Physics and Astronomy, Texas A$\&$M
University-Commerce, Commerce, TX 75429-3011, USA}
\author{Bao-An Li\footnote{Corresponding author: Bao-An.Li$@$tamuc.edu}}
\affiliation{Department of Physics and Astronomy, Texas A$\&$M
University-Commerce, Commerce, TX 75429-3011, USA}
\author{Lie-Wen Chen\footnote{%
Email: lwchen$@$sjtu.edu.cn}} \affiliation{Department of Physics and
Astronomy and Shanghai Key Laboratory for Particle Physics and
Cosmology, Shanghai Jiao Tong University, Shanghai 200240, China}
\affiliation{Center of Theoretical Nuclear Physics, National
Laboratory of Heavy Ion Accelerator, Lanzhou 730000, China}

\date{\today}

\begin{abstract}
Neutron-skins in coordinate and proton-skins in momentum are predicted to coexist in heavy nuclei and their correlation is governed by
Liouville's theorem and Heisenberg's uncertainty principle. An analysis of their correlation within a further extended Thomas-Fermi
(ETF$^+$) approximation incorporating effects of nucleon short-range correlations reveals generally that protons move faster than neutrons in neutron-skins of heavy
nuclei.

\end{abstract}

 \maketitle

\textit{1. Introduction:} Understanding the neutron-proton (isospin)
asymmetry dependence of nuclear equation of state (EOS) and the
underlying isovector strong interactions is a longstanding and
common goal of contemporary nuclear physics and astrophysics.
Various radioactive beam facilities being built around the world,
electron, hadron, light- and heavy-ion beams available from low to
high energies, advanced x-ray satellites and gravitational wave
detectors in operation together provide multiple tools for realizing ultimately
the stated goal. In fact, many observables and
phenomena in both terrestrial nuclear experiments and astrophysical
observations have been used to probe the poorly known nature of
neutron-rich nucleonic matter especially the symmetry energy
term of its EOS \cite{EPJA}. In particular, recognizing that neutron-skins of heavy nuclei
provide a great testing ground of isovector interactions,
much efforts have been devoted to measuring the sizes of neutron-skins
using many methods ranging from photopion production, pionic and antiprontic atoms,
hadron-nucleus scatterings to parity-violating electron scatterings, see, e.g., refs. \cite{Vin14,Tamii,Hor14} for recent reviews.
While the community has yet to reach a consensus on the precise values of neutron-skins of heavy nuclei,
the studies have been extremely fruitful. On the other hand, it is well known both theoretically \cite{Mig57,Lut60,Bethe,Ant88,Pan97,Mah92}
and experimentally \cite{SRC,Sub08,Hen14} that short-range nucleon-nucleon correlations (SRC) due to the tensor
components and/or the repulsive core of nuclear forces lead to the formation of a high-momentum tail (HMT)
in the single-nucleon momentum distribution, see, e.g., refs. \cite{Arr12,Cio15} for recent reviews. Analyses of a few recent experiments have revealed some possible indications of the isospin-dependence of the SRC.
For example, nucleon spectroscopic factors extracted from knock-out reactions induced by radioactive beams~\cite{Gade}, proton occupations from
dispersive optical model analyses of proton scattering on Ca isotopes~\cite{Bob06} and systematic studies of triple coincidence measurements of $(e,e'pn)$ and $(p,p'pn)$ reactions \cite{Sub08,Hen14}
all indicate consistently that the minority particle in isospin-asymmetric nuclei is relatively more correlated. Moreover, calculations based on the Variational Monte Carlo (VMC) \cite{VMC} and the neutron-proton dominance model ~\cite{Hen14} have shown that the average kinetic energy of protons is significantly higher than that of neutrons in neutron-rich nuclei from $^{8}$He to $^{208}$Pb.
In this work, incorporating the SRC effects in an extended Thomas-Fermi model, we show for the first time that on average protons move faster than neutrons
in neutron-skins of heavy nuclei. Proton-skins in momentum (k-) space coexist with neutron-skins in coordinate (r-) space in heavy nuclei and their correlation is
governed by Liouville's theorem and Heisenberg's uncertainty principle.

\vspace*{0.5cm}

\textit{2. SRC-Modified Single-Nucleon Momentum Distribution
Function $n_{\v{k}}^J$ in Isospin-Asymmetric Nucleonic Matter:}
Guided by earlier findings from analyses of both experimental results \cite{Hen14,Hen15b,Hen15} and microscopic many-body calculations, see, e.g., refs. \cite{mu04,Rio09,Rio14,Yin13,ZHLi},
we parameterize the single-nucleon momentum distribution $n_{\v{k}}^J$ in isospin asymmetric and cold nucleonic matter \cite{Hen14,Cai15,Cai16,Cai16a} with
\begin{equation}\label{MDGen}
n^J_{\v{k}}(\rho,\delta)=\left\{\begin{array}{ll}
\Delta_J,~~&0<|\v{k}|<k_{\rm{F}}^J,\\
&\\
\displaystyle{C}_J\left({k_{\rm{F}}^{J}}/{|\v{k}|}\right)^4,~~&k_{\rm{F}}^J<|\v{k}|<\phi_Jk_{\rm{F}}^J
\end{array}\right.
\end{equation}
where $k_{\rm{F}}^J$ is the Fermi momentum of the nucleon J.  The
$\Delta_J$ measures the depletion of the Fermi sea with respect to
the step function for a free Fermi gas (FFG). The three parameters
$\Delta_J$, $C_J$ and $\phi_J$ depend linearly on the isospin
asymmetry $\delta\equiv (\rho_{\rm{n}}-\rho_{\rm{p}})/\rho$ in a
general form of $Y_J=Y_0(1+Y_1\tau_3^J\delta)$ where
$\tau_3^{\rm{n}}=+1$ and $\tau_3^{\rm{p}}=-1$
\cite{mu04,Rio09,Rio14,ZHLi,Yin13}. We notice that this
parameterization reduces exactly to the one used in ref.
\cite{Hen15b} in the limiting case of symmetric nuclear matter (SNM). The amplitude ${C}_J$ and
the high-momentum cutoff coefficient $\phi_J$ determine the fraction
of nucleons in the HMT via $
x_J^{\rm{HMT}}=3C_{{J}}\left(1-\phi_J^{-1}\right)$. The latter varies approximately linearly with $\delta$  consistent with 
earlier predictions \cite{mu04,Rio09,Rio14,Yin13}.
The normalization condition $[{2}/{(2\pi)^3}]\int_0^{\infty}n^J_{\v{k}}(\rho,\delta)\d\v{k}={(k_{\rm{F}}^{J})^3}/{3\pi^2}
$ requires that only two of the three parameters are independent.
We emphasize that our parameterization is also constrained by the EOS of pure neutron matter (PNM) obtained from
microscopic many-body theories
\cite{Sch05,Epe09a,Gez10,Stew10,Tew13,Gez13}. In particular, the
contact $C_{\rm{n}}^{\rm{PNM}}$ for PNM is obtained by applying Tan's adiabatic sweep theorem \cite{Tan08} to the EOS of PNM, see
ref.\,\cite{Cai15} for more details.

\begin{figure}[h!]
\centering
  \includegraphics[width=8.5cm]{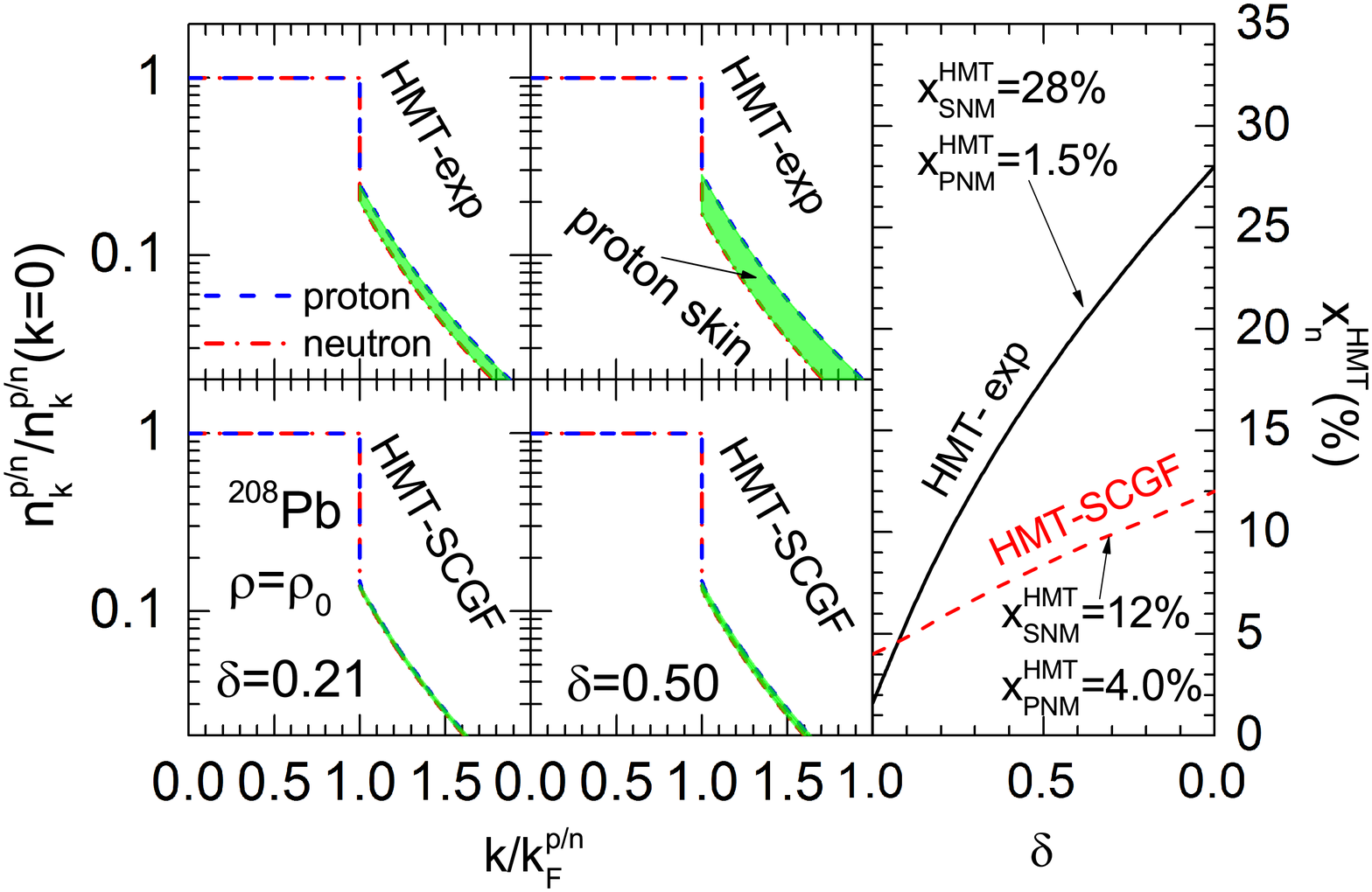}
  \caption{(Color Online). Reduced nucleon  momentum distribution (normalized to 1 at zero momentum) of neutron-rich nucleonic matter with an isospin asymmetry of $\delta=0.21$ (left) and 0.50 (middle) reachable respectively in the core and surface of $^{208}$Pb, and the fraction of neutrons in the high momentum tail as a function of $\delta$ using the HMT-exp and HMT-SCGF parameter set (right), respectively.}
  \label{fig_xhmt}
\end{figure}

Systematic analyses of many experiments, see e.g., refs.
\cite{Hen14,Hen15b,Hen15}, indicate that the percentage of nucleons
in the HMT is about 25$\%$ in SNM.  While
various many-body theories have consistently predicted the SRC effects on the HMT qualitatively consistent with the experimental
findings, the predicted size of the HMT still depends on the model and
interaction used. For example,  the Self-Consistent Green's Function
(SCGF) theory using the Av18 interaction predicts a 11-13\% HMT for
SNM at saturation density $\rho_0$ \cite{Rio09,Rio14}. While the latest Bruckner-Hartree-Fock calculations
predict a HMT ranging from about 10\% using the N3LO450 to
over 20\% using the Av18, Paris or Nij93 interactions \cite{ZHLi},
the latest VMC calculations for $^{12}$C gives a 21\% HMT
\cite{VMC}. Little information about the isospin dependence of the HMT has been extracted from experiments so far. While based on the observation that 
the SRC strength of a neutron-proton pair is about 18-20 times that of two protons, the HMT in PNM was estimated to be about 1-2\% \cite{Hen15b}.  However, some recent calculations indicate a significantly higher HMT in PNM. For example, the SCGF theory predicted a 4-5\% HMT in PNM \cite{Rio09,Rio14}.

Recognizing the aforementioned discrepancies and still model dependent predictions about the size of the HMT, we carried out extensive calculations by varying the size of HMT in both SNM and PNM. 
We present here results using two sets of model parameters leading to $x_J^{\rm{HMT}}$ values resembling those from the experimental analysis \cite{Hen14} and the SCGF predictions \cite{Rio09,Rio14}, respectively. 
More quantitatively, with the HMT-exp parameter set of $x_{\rm{SNM}}^{\rm{HMT}}=28\%,~~
x_{\rm{PNM}}^{\rm{HMT}}=1.5\%$, we have
$C_0=0.161$, $C_1=-0.25$,
$\phi_0=2.38$ and $\phi_1=-0.56$. While
with the HMT-SCGF parameter set of $ x_{\rm{SNM}}^{\rm{HMT}}=12\%$ and
$x_{\rm{PNM}}^{\rm{HMT}}=4\% $ we have $\phi_0=1.49,
\phi_1=-0.25$, $C_0=0.121$ and $C_1=-0.01$. The resulting neutron
fractions in the HMT as a function of $\delta$ at $\rho_0$
are shown in the right window of Fig. \ref{fig_xhmt} for both parameter sets. The reduced nucleon momentum distributions
(normalized to 1 at zero momentum) with $\delta=0.21$ and $0.5$ for both cases are shown on
the left. While the isospin asymmetry $\delta=0.21$ can be easily realized in the core, $\delta=0.5$ can be reached in the surface area
of $^{208}$Pb. Clearly, relative to the center in k-space, nucleonic matter has a distinct proton-skin and its thickness grows with the isospin asymmetry at a rate depending on the sizes of the HMT parameters used.

The $1/k^4$ behavior of the HMT and its cut-off parameter $\phi_0$ were first introduced by Hen et al in \cite{Hen15} for SNM based on the VMC predictions for the deuteron momentum distribution and the finding from many calculations that the HMT of other nuclei is {\it approximately} proportional to that of deuteron. The fundamental origin for these observations is the dominating tensor force in the spin-triplet (S=1) and isospin-singlet (T=0) neutron-proton interaction channel. Our extension of the high-momentum cut-off parameter $\phi_J$ to include an isospin-dependent term $\phi_1\tau_3\delta$ for asymmetric nuclear matter is numerically consistent with existing experimental constraints on the HMT and necessary to extrapolate meaningfully from SNM to PNM where it was predicted to have an approximately1-5\% HMT due to probably the repulsive core instead of the tensor force. Numerically, the  magnitude of the $\phi_1\tau_3\delta$ is only about 5\% and 12\% of the leading constant term in $\phi_J$ for $^{208}$Pb with the HMT-SCGF and HMT-exp parameters, respectively. First of all, the currently estimated uncertainty range of $\phi_0$ \cite{Hen15b,Hen15} can tolate such a weakly isospin-dependent term. Secondly, the weak isospin-dependence of the cut-off parameter is consistent with model predictions for single-nucleon momentum distributions in asymmetric matter \cite{Rio14,Yin13}. Moreover, as emphasized very recently in ref. \cite{Mas16}, the dominance of the (S=1,T=0) neutron-proton paris over other nucleon pairs in the HMT can now be more accurately quantified in advanced and realistic calculations. In particular, it was pointed out that {\it ``states different from the deuteron one, namely the states (01) and (11), do contribute to the high momentum part of the momentum distributions, demonstrating, in the case of the state (11), that a considerable number of two-nucleon states with odd value of the relative orbital momentum is present in the realistic ground-state wave function of nuclei''.} Experimentally, it was shown that there is an approximately 5-15\% constant contribution to the HMT from pp (nn) pairs probably due to the isoscalar repulsive core of nuclear interactions \cite{SRC}. Thus, if one considers contributions of all nucleon pairs, one expects that the parameters characterizing the HMT to be isospin dependent. In this work, we have assumed that the $1/k^4$ behavior of the HMT also works in isospin asymmetric nuclear matter. The small HMT in PNM is then accounted for consistently within our framework by using the weakly isospin dependent cut-off parameter $\phi_J$.

\vspace*{0.5cm}
\textit{3. Further Extended Thomas-Fermi (ETF$^+$) Approximation Incorporating SRC Effects:} In the
original ETF framework which is a semi-classical approximation to the Hartree-Fock theory, the nucleon kinetic energy density profile in finite nuclei
\begin{equation}\label{epskin1}
\varepsilon_J^{\rm{kin}}(r)=\frac{1}{2M}\left[\alpha_J^{\infty}\cdot\rho_J^{5/3}(r)+\frac{\eta_J}{36}\frac{(\nabla\rho_J(r))^2}{\rho_J(r)}+\frac{1}{3}\Delta\rho_J(r)\right]
\end{equation}
was obtained by truncating the Wigner-Kirkwood expansion of the
Block density matrix at the order of
$\hbar^2$\,\cite{Bra85,Kri79,Cam80,Bar83}.
The first term originally with $\alpha_J^{\infty}=(3/5)(3\pi^2)^{2/3}$ is the bulk part as if
nucleons are in infinite nuclear matter and have a step function for
their momentum distributions. The second term originally proposed by Weizs\"acker \cite{Bra85,Wei35} is very
sensitive to surface properties of finite nuclei. Its strength factor $\eta_J$ has been under debate \cite{Bra85} and was 
found to affect significantly the halo and/or skin nature of the surfaces of heavy nuclei \cite{Luk15}. 
The last term involving a Laplacian operator is normally very small
as the nuclear surfaces are generally very smooth.
We emphasize that the above relationship is general regardless how the density profile is obtained. 
In the following, we refer our calculations considering the SRC effects as the ETF$^+$ to
distinguish it from the original ETF.

First, we discuss how the HMT affects the bulk part of the kinetic energy. 
With the SRC-modified single-nucleon momentum distribution function of Eq. (\ref{MDGen}), the kinetic energy density in infinite matter is
given by
\begin{equation}\label{epskin}
\frac{2}{(2\pi)^3}\int_0^{\phi_Jk_{\rm{F}}^J}\frac{\v{k}^2}{2M}n_{\v{k}}^J\d\v{k}=\frac{1}{2M}\frac{3}{5}(3\pi^2)^{2/3}\rho_J^{5/3}\Phi_J
\end{equation}
where $\Phi_J=1+C_J(5\phi_J+{3}/{\phi_J}-8)>1$ is determined by
properties of the HMT. Thus, the original $\alpha_J^{\infty}$ in Eq.
(\ref{epskin1}) is enhanced by the SRC factor $\Phi_J$ to
$\alpha_J^{\infty}=(3/5)(3\pi^2)^{2/3}\Phi_J$. In neutron-rich
systems, since relatively more protons are depleted from the Fermi
sea to form a proton-skin in the HMT, the bulk part of the kinetic
energy density is enhanced more for protons than neutrons. For the
HMT-exp parameter set, we find $\Phi_{\rm{p}}=2.09$
and $\Phi_{\rm{n}}=1.60$ for isospin asymmetric
matter with $\delta=0.21$. While for the HMT-SCGF, we have
$\Phi_{\rm{p}}=1.21$ and $\Phi_{\rm{n}}=1.14$.

Similar to the measure of the neutron-skin in r-space, $\Delta r_{\rm{np}}\equiv\langle
r_{\rm{n}}^2\rangle^{1/2}-\langle r_{\rm{p}}^2\rangle^{1/2}$ with
$\langle r_{\rm{n/p}}^2\rangle^{1/2}$ the RMS radius of neutrons/protons, one may quantify the proton-skin in k-space for finite nuclei using the difference between the average kinetic energies of protons and neutrons, i.e.,
$\Delta E_{\rm{pn}}^{\rm{kin}}\equiv\langle E_{\rm{p}}^{\rm{kin}}\rangle-\langle E_{\rm{n}}^{\rm{kin}}\rangle$, with
\begin{equation}\label{EkinJ}
\langle E_J^{\rm{kin}}\rangle=\left.\int_0^{\infty}\varepsilon_J^{\rm{kin}}(r)\d\v{r}\right/{\int_0^{\infty}
\rho_J(r)\d\v{r}}\equiv \langle k_J^2\rangle/2M
\end{equation}
where $M$ is the average mass of nucleons and $\langle k_J^2\rangle$ is the RMS radius squared in k-space for the nucleon J.
To evaluate surface properties, one has to specify the nucleon's density profiles $\rho_J(r)$.
We adopt here the 2-parameters Fermi (2pF) distribution widely used in the literature, i.e.,
$\rho_J(r)=\rho_0^J[1+\exp(({r-c_J})/{a_J})]^{-1}$, where $c_J$ and
$a_J$ are the half-density radius and diffuseness parameter,
respectively. While our formalism and conclusions are general, in the following we use
$^{208}\rm{Pb}$ as an example for numerical calculations.
While the $a_{\rm{p}}$ and $c_{\rm{p}}$ of $^{208}\rm{Pb}$ are
constrained by experiments to $a_{\rm{p}}\approx0.447\,\rm{fm}$ and
$c_{\rm{p}}\approx6.680\,\rm{fm}$\,\cite{Jon14}, the corresponding values for neutrons are still poorly known. We
explore the correlation between the two kinds of skins in the range of $0.01 \leq \Delta r_{\rm{np}}\leq 0.43$\,fm
by taking $a_{\rm{n}}=0.55\pm0.05\,\rm{fm}$ and $c_{\rm{n}}=6.8\pm0.2\,\rm{fm}$ \cite{Vin14}, while the fiducial value
of $\Delta r_{\rm{np}}\approx0.159\,\rm{fm}$ \cite{Dan14} is used for some illustrations. 

Next we explain how the surface strength factor $\eta_J$ is constrained using as much as possible experimental information. 
Since the proton density profile is experimentally known, the
$\eta_{\rm{p}}$ is uniquely determined for a given $\langle
E_{\rm{p}}^{\rm{kin}}\rangle$. For neutrons, however, there are
three unknowns $a_{\rm{n}}$, $c_{\rm{n}}$ and $\eta_{\rm{n}}$ as the
neutron density profile is not precisely known. Thus, given an
average kinetic energy $\langle E_{\rm{n}}^{\rm{kin}}\rangle$ of
neutrons,  for any specific value of neutron-skin $\Delta
r_{\rm{np}}$ only a correlation among the $a_{\rm{n}}$, $c_{\rm{n}}$ and $\eta_{\rm{n}}$ is constrained.  For the HMT-exp parameter set, we use  $\langle
E_{\rm{p}}^{\rm{kin}}\rangle\approx41.9\,\rm{MeV}$ and $\langle
E_{\rm{n}}^{\rm{kin}}\rangle\approx34.0\,\rm{MeV}$ extracted by Hen
et al. for $^{208}\rm{Pb}$ in ref.\,\cite{Hen14} using the
neutron-proton dominance model with its parameters constrained by
their experimental data. With about a factor of two smaller (larger) fractions of HMT nucleons in SNM (PNM) in the HMT-SCGF parameter set, 
the nucleon average kinetic energies are expected to be smaller at moderate $\delta$ values. While the SCGF theory has predicted the average nucleon kinetic energies for infinite matter, 
to our best knowledge, no prediction for $^{208}$Pb is currently available. To compare calculations using different HMT parameters, we set $\langle E_{\rm{p}}^{\rm{kin}}\rangle=\langle
E_{\rm{n}}^{\rm{kin}}\rangle$ in the HMT-SCGF calculations and vary their values within a band of 2 MeV around $34.0\,\rm{MeV}$. The latter is the value of $\langle
E_{\rm{n}}^{\rm{kin}}\rangle$ used in the HMT-exp calulcations.
As an example, using $\Delta r_{\rm{np}}\approx0.159\,\rm{fm}$, for the HMT-exp parameter set we found
$\eta_{\rm{p}}\approx26.7$ and $\eta_{\rm{n}}=7.2\sim9.2$,
respectively. While for the HMT-SCGF, we have $\eta_{\rm{p}}\approx
45.9$ and $\eta_{\rm{n}}=41.5\sim 46.5$, respectively. It is
interesting to note that the HMT-SCGF case requires a much large
surface contribution to reproduce the same nucleon kinetic energies
used in the HMT-exp calculations. This is understandable as the bulk part of the kinetic energy in the HMT-SCGF calculations is not enhanced by the HMT as much as in the HMT-exp calculations.

\vspace*{0.5cm}
\begin{figure}[h!]
\centering
  \includegraphics[width=8.5cm]{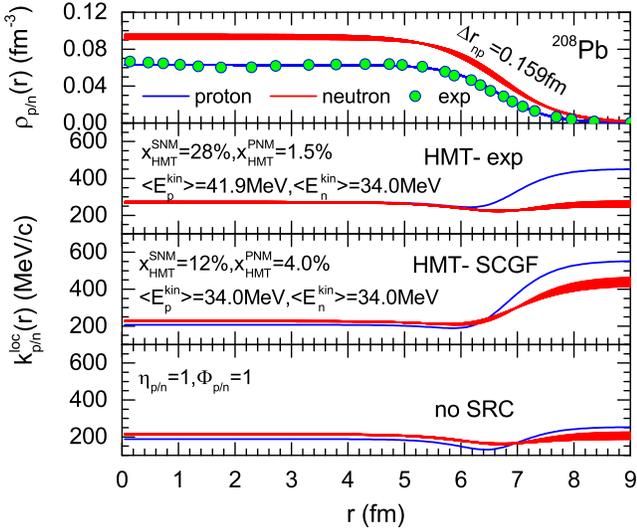}
  \caption{(Color Online). Input local density (upper) and calculated momentum profiles (lower 3 windows) in $^{208}\rm{Pb}$ using the parameters specified.}
  \label{fig_RKdis_ex}
\end{figure}
\textit{4. Coexistence of Neutron-Skins in r-Space and Proton-Skins in k-Space in Heavy Nuclei:}
A key quantity for our discussions here is nucleons'
average local momentum $k^{\rm{loc}}_J(r)$ defined through the local kinetic energy per nucleon
\begin{equation}\label{kloc}
\langle E_J^{\rm{loc}}(r)\rangle=\varepsilon_J^{\rm{kin}}(r)/\rho_J(r)\equiv\left(k^{\rm{loc}}_J(r)\right)^2/2M.
\end{equation}
In Fig. \ref{fig_RKdis_ex}, we examine the correlation between the
neutron-skin in r-space and proton-skin in k-space in heavy nuclei
by comparing nucleons' average local density and momentum as a
function of radius $r$. First of all, as a reference, nucleon local momenta
in the original EFT calculations without considering the SRC effects
are shown in the bottom window. The width of the band reflects the uncertainties of the input quantities. In the interior, neutrons
have higher local momenta due to their higher densities than
protons. In the surface area, very interestingly, because
protons have larger values of the Weizs\"acker surface term
$(\nabla\rho_J/\rho_J)^2$ they have higher local momenta than
neutrons, indicating the coexistence of a proton-skin in k-space and
a neutron-skin in r-space. Analytically, we have approximately in
the outer surface area
$k_{\rm{p/n}}(r)\approx 1/(72M)(\nabla\rho_{\rm{p/n}}/\rho_{\rm{p/n}})^2\approx 1/(72Ma_{\rm{p/n}}^2)$,
leading to $k_{\rm{p}}^{\rm{loc}}(r)>k_{\rm{n}}^{\rm{loc}}(r)$ since
the protons' surface diffuseness $a_{\rm{p}}$ is normally much less
than the $a_{\rm{n}}$ for neutrons in heavy nuclei. Turning on the
SRC effects with either the HMT-exp or HMT-SCGF parameters, most
interestingly, protons have much larger local momenta than neutrons in the surface area. Moreover, because of the stronger surface contributions in calculations with the HMT-SCGF
parameter set, the local momenta of both neutrons and protons in the surface area are
higher than those in the HMT-exp calculations. We also found that calculations with the HMT-SCGF
parameters by varying the $\langle E_{\rm{p}}^{\rm{kin}}\rangle=34$\,MeV within a 2\,MeV range, or
the size of neutron-skin around $\Delta r_{\rm{np}}=0.159$ fm within a large range of about 0.15 fm do not change the qualitative
features of our results. As we shall discuss next, the observed coexistence of proton-skins in k-space and neutron-skins in r-space in heavy nuclei is a requirement
of quantum mechanics.

\vspace*{0.5cm}
\begin{figure}[h!]
\centering
  \includegraphics[width=8.5cm]{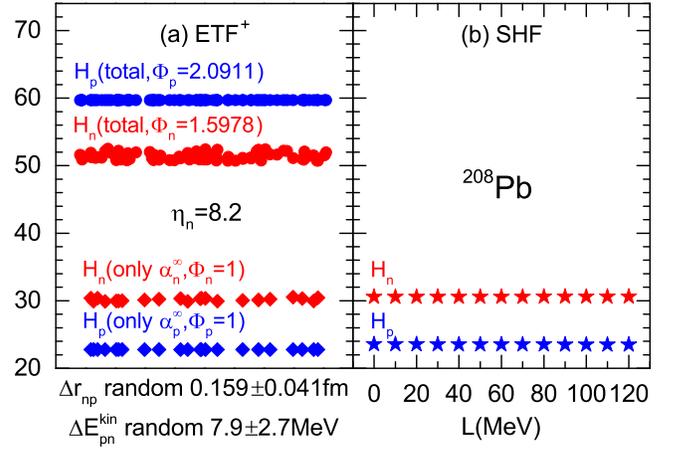}  \caption{(Color Online). $H_J$ obtained from the ETF$^+$ calculations using the HMT-exp parameter set and the SHF prediction.}
  \label{fig_Hr2k2}
\end{figure}
\textit{5. Inverse Correlation between the RMS Radii in r- and k-Spaces:}
Defining $H_J\equiv\langle
r_J^2\rangle\langle k_J^2\rangle$, we first present general arguments then numerical results illustrating that
$H_J$ is a constant in a given model. First of all, Heisenberg's uncertainty principle sets a limit on the $\delta
r_J\delta k_J\gtrsim 1$ with $\delta r_J$ and $\delta k_J$ the
standard deviations in the r- and k-space, respectively. Moreover, Liouville's theorem requires that the phase
space density $f_J(\v{r},\v{k})$ is a constant of motion\,\cite{Lan80St}. Thus,  one expects that
$2\times4\pi\langle r_J\rangle^3/3\times4\pi\langle
k_J\rangle^3/3/(2\pi)^3=N_J$ with $N_{\rm{n}}=N$ and $N_{\rm{p}}=Z$,
where $\langle r_J\rangle$ and $\langle k_J\rangle$ are the effective
radii of the r- and k-space, respectively. Consequently,
$\langle r_J\rangle\langle k_J\rangle=(9\pi N_J/4)^{1/3}$ is a
constant only depending on the number $N_J$. Putting these constraints together, we then expect that
$\langle r_J^2\rangle \langle k_J^2\rangle=\langle r_J\rangle^2\langle
k_J\rangle^2[ 1+({\delta r_J}/{\langle r_J\rangle})^2 +({\delta
k_J}/{\langle k_J\rangle})^2]+(\delta r_J)^2(\delta
k_J)^2\approx\rm{constant}$ if $\delta r_J\lesssim\langle r_J\rangle$
and $\delta k_J\lesssim\langle k_J\rangle$. Indeed, the latter conditions are well satisfied in heavy nuclei.
For instance, applying the $n^J_{\v{k}}(\rho,\delta)$ of Eq. \ref{MDGen} for nucleons in SNM with the HMT-exp parameters,
$(\delta k/\langle k\rangle)^2\approx0.03$, and using the 2pF distribution for neutrons in $^{208}$Pb with $\Delta r_{\rm{np}}=0.159$ fm,  $(\delta r_{\rm{n}}/\langle
r_{\rm{n}}\rangle)^2\approx0.09$, which are both negligibly small.
In the extreme case of a uniform phase space density, i.e.,
$f_J(\v{r},\v{k})=\Theta(R_J-|\v{r}|)\Theta(K_J-|\v{k}|)$, where
$R_J$ and $K_J$ are the hard-sphere radii in r- and k-space, respectively, and $\Theta$ is the step
function, $\langle r_J^2\rangle\langle
k_J^2\rangle=(4/25\pi)(9\pi/4)^{5/3}N_J^{2/3}\equiv H_J^0$.

We now turn to numerically testing the constancy of $H_J$.
As an example, we use the HMT-exp parameter set and randomly select with equal weights the following quantities within their respective uncertainty ranges around their central values,
 $\Delta r_{\rm{np}}\approx0.159\pm0.041\,\rm{fm}$, $\langle
E_{\rm{p}}^{\rm{kin}}\rangle\approx41.9\pm2.3\,\rm{MeV}$ and
$\langle
E_{\rm{n}}^{\rm{kin}}\rangle\approx34.0\pm1.5\,\rm{MeV}$ (corresponding to $\Delta E_{\rm{pn}}^{\rm{kin}}\approx7.9\pm2.7\,\rm{MeV}$). The
results are shown in the left window of Fig. \ref{fig_Hr2k2}. The
lower two chains of  red and blue diamonds denoted by
``$H_{\rm{n}}\,(\rm{only\,}\alpha_{\rm{n}}^{\infty},\Phi_{\rm{n}}=1)$"
and
``$H_{\rm{p}}\,(\rm{only\,}\alpha_{\rm{p}}^{\infty},\Phi_{\rm{p}}=1)$", respectively, 
are results of the original ETF model using neither the surface terms nor the
SRC effects, i.e,  $\eta_{\rm{p/n}}=0$ and no Laplacian term.
The upper two chains are from the full ETF$^+$ model calculations with
$\eta_{\rm{n}}=8.2$. For a comparison, shown in the right window are results of the Skyrme-Hartree-Fock
(SHF) calculations by varying the  slope parameter
$L\equiv L(\rho_0)$ of the symmetry energy at $\rho_0$ from 0 to 120\,MeV using the MSL0 parameter
set\,\cite{Che10}. Since the SHF does not contain SRC effects, its
predictions are closer to the original ETF calculations \cite{Cas87}. Interestingly, as expected based on basic principles of quantum mechanics, in all models
considered the $H_{\rm{n}}$ and $H_{\rm{p}}$ are essentially all constants. The most important consequence is that as nucleons' RMS radius in k-space is increased by the HMT, their RMS radius 
in r-space has to decrease correspondingly. This leads to the coexistence of and a positive correlation between the neutron-skin in r-space and proton-skin in k-space of heavy nuclei. 

\begin{figure}[h!]
\centering
  \includegraphics[width=8.5cm]{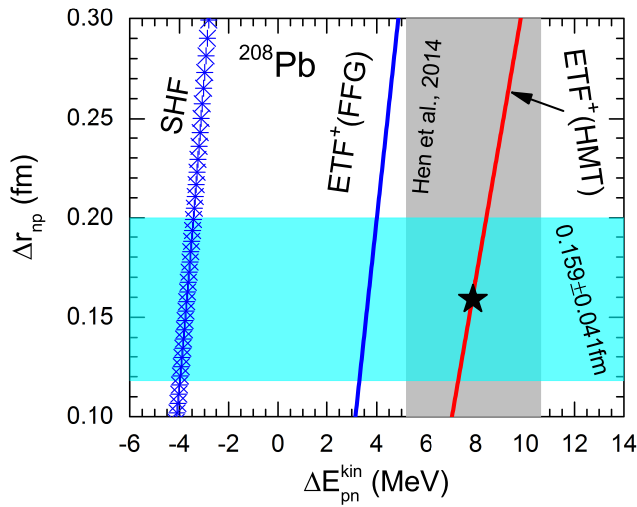}
  \caption{(Color Online). Correlation between the neutron-skin in r-space $\Delta r_{\rm{np}}$ and proton-skin in k-space $\Delta E_{\rm{pn}}^{\rm{kin}}$ for $^{208}\rm{Pb}$ within the ETF$^+$ approach with the SRC effect (HMT) using the HMT-exp parameter set or without it (FFG).  The black star represents the central cross point of the constraints on both the $\Delta r_{\rm{np}}$ and $\Delta E_{\rm{pn}}^{\rm{kin}}$.}
  \label{fig_diffEkinSkinCorr}
\end{figure}
\textit{6. Combined Proton-Skin and Neutron-Skin Constraints on Nuclear Models:} 
Shown in Fig. \ref{fig_diffEkinSkinCorr} are correlations between the sizes of neutron-skin $\Delta
r_{\rm{np}}$ in r-space and proton-skin $\Delta E_{\rm{pn}}^{\rm{kin}}$ in k-space for $^{208}\rm{Pb}$ within the ETF$^+$ approach incorporating the SRC effects (HMT) using the HMT-exp parameter set or without considering them (FFG). Also shown are predictions by the SHF approach. Several important observations can be made. Firstly, as one expects, predictions of the SHF model and the
ETF$^+$(FFG) without considering the SRC effects do not satisfy simultaneously the combined constraints on the sizes of both the neutron-skin and proton-skin. Secondly, the sizes of neutron-skin $\Delta r_{\rm{np}}$ and proton-skin $\Delta E_{\rm{pn}}^{\rm{kin}}$ are strongly correlated approximately linearly within their existing constraints. Thus, measuring more accurately the size of either the neutron-skin in r-space or proton-skin in k-space will help improve our knowledge about the same physics underlying both quantities. 

\vspace*{0.5cm}
\textit{7. Conclusion:} In conclusion, protons move faster than neutrons in neutron-skins of heavy nuclei. The neutron-skins in r-space and proton-skins in k-space coexist and they are intrinsically correlated as required by Liouville's theorem and Heisenberg's uncertainty principle. A precise measurement of either one of them will help constrain the other one and improve our knowledge about nuclear surfaces in the complete phase space. 

 \vspace*{0.5cm}
\textit{Acknowledgement:} We would like to thank O. Hen, W.G. Newton and Z. Taylor for helpful discussions. This work is supported in part by the U.S. Department of Energy Office of Science under Award Number DE-SC0013702, the CUSTIPEN (China-U.S. Theory Institute for Physics with Exotic Nuclei) funded by the U.S. Department of Energy, Office of Science under grant number DE-SC0009971, the Major State Basic Research Development Program (973 Program) in China under Contract Nos. 2015CB856904 and 2013CB834405, the NSFC under Grant Nos. 11320101004,11275125, 11135011 and 11625521, the \"Shu Guang" project supported by Shanghai Municipal Education Commission and Shanghai Education Development Foundation, the Program for Professor of Special Appointment (Eastern Scholar) at Shanghai Institutions of Higher Learning, and the Science and Technology Commission of Shanghai Municipality (11DZ2260700). We would also like to thank the Texas Advanced Computing Center (TACC) for proving computing resources for this work.

\end{document}